# Improving the electronic and optical properties of Carbz-PAHTDDT-based dyes through chemical modifications

Narges Mohammadi[*] and Feng Wang[*]

eChemistry Laboratory, Faculty of Life and Social Sciences, Swinburne University of Technology, Hawthorn, Melbourne, Victoria, 3122, Australia

* Corresponding authors:

nmohammadi@swin.edu.au (Tel.: +61 3 9214 8785 N. Mohammadi).

fwang@swin.edu.au (Tel.: +61 3 9214 5065 F. Wang).

**Abstract**

To investigate geometric and electronic structure, a theoretical study is performed on the Carbz-PAHTDDT (S9) organic dye sensitizer. This dye has a reported promising efficiency when coupled with ferrocene-based electrolyte composition. The present study indicated that the long-range correction to the theoretical model in the time-dependent density functional theory is important to produce accurate absorption wavelengths. In the present study, the chemical structure of the original Carbz-PAHTDDT dye on the π-conjugated bridge is also rationally changed to produce new dyes aiming at enhancing the spectral response as a desirable property of organic dyes in DSSC application. The theoretical studies on the new dyes have shown a significant red-shifting and broadening of their absorption spectra.

**Keywords:** rational dye design; DSSC; HOMO-LUMO gap; BHandH functional; Organic dye sensitizer; Carbz-PAHTDDT dye; red shift;

## 1. Introduction

With a booming in research effort to develop cost-effective renewable energy devices, dye sensitized solar cells (DSSC), also known as Grätzel cells [1] have been the topic of more than a thousand published papers just in 2010 [2], yet without a significant progress in the device performance in recent years. A 7% efficient DSSC was reported in the seminal Nature paper of Grätzel and O'Regan published in 1991 [1]. Three years later, Grätzel group could achieve an efficiency of 10% [3]. In spite of the immense research effort to enhance the efficiency of DSSC afterwards, not a large increase in efficiency could be achieved. Until recently, the highest solar to electricity energy conversion efficiencies exceeding 11% belonged to cells using ruthenium-based dye photosensitizers N3 [3, 4], N719 [4-6] and black dye [7-9], together with titanium dioxide semiconductor and iodide/triiodide redox couple [4, 5, 9, 10]. To achieve this efficiency, internal energy levels of all of the three main components of DSSC (i.e. semiconductor, dye sensitizer, and redox shuttle) have been well tuned [11].

A couple of main barriers prevent the DSSCs from rapidly improving the efficiency over the past several years. First, many of the studies in DSSCs have focused on improving only one of the main components of the device at a time which usually results in deviation from the fine-tuned energy levels of DSSC's standard components. As suggested by Hamann *et al.* [11], it needs at least two of the three main components of DSSC to be altered simultaneously to overcome the ~11% efficiency plateau. A recent study of Yella *et al.* published in Science [12], in which the highest DSSC efficiency to this date exceeding 12% is reported, gives a good example to address this issue. In their study, the impressively high efficiency was gained by incorporating a cobalt-based redox mediator replacing the iodide/triiodide redox couple in conjugation with a porphyrin-based dye which was specifically designed to retard interfacial back electron transfer and was co-sensitized with another organic dye sensitizer to improve the light-harvesting efficiency. In the same manner, Daeneke *et al.* [13] reported a highly efficient DSSC based on ferrocene electrolyte. In their study, ferrocene/ferrocenium (Fc/Fc$^+$) redox couple was utilized and was coupled with a novel organic dye sensitizer called Carbz-PAHTDTT (S9) dye. Previous attempts to replace the conventional iodide/triiodide with ferrocene-based redox couples led to very low efficiencies ($\eta$<0.4%) [14, 15], where changes have only been made on a single component (i.e. redox couple); whereas the dye sensitizer, i.e. conventional N3 dye remained unchanged. As a result, it is important to carefully configure the energy levels of different components of DSSC when altering the standard components.

Another obstacle contributing to the stagnation of the efficiency of DSSC is the bottleneck of the design and test of the new materials (e.g. dye sensitizer) for DSSCs, which have been dominated by the often costly and time-consuming synthesis procedures [16]. As in the case of new dye sensitizer materials development, it is difficult for synthetic chemists to work out high-performance dyes with the desirable properties prior to the experiments on the assembled cell, without any support on the information of the new dyes [17]. For example, the energy conversion efficiencies of the recently constructed DSSCs based on two chemically similar dyes were very different [18]. One is $\eta$=6.79% and the other is $\eta$=4.92%. And interestingly, the two dyes only differ in their π-spacers: one takes thiophene ($\eta$=6.79%) and the other is thiazole ($\eta$=4.92%). Both spacers have a sulphur embedded in the pentagon ring, but the former contains two C=C bonds and the latter has one C=C bond and one C=N bond. Such the structure and property relationship of the new dyes is hardly obtained from "chemical intuitions" without accurate quantum mechanical calculations.

In some cases, disappointing results from the most late-stage of the dye synthesis laboratories indicate an urgent need to understand the physical origin of dyes at molecular level, prior to experiments taking place. To overcome this bottleneck in the development of new DCCSs with better efficiency, the state-of-art computational methods need to be utilised. Today, accurate first-principle quantum chemical calculations are made available on supercomputing facilities accessible to more research groups. Such calculations are a reliable tool to design, study, and screen new materials prior to synthesis. Computer-aided rational design of new dye sensitizers based on the systematic chemical modifications of the dye structure has recently drawn the attention of several groups, including ours [19-27].

The recent breakthrough of the DSSC based on $Fc/Fc^+$ redox couple and the Carbz-PAHTDTT (S9) dye sensitizer [13], stimulated the present study with more theoretical insight to obtain desirable properties for the new dyes through rational chemical modifications of the original dye. To the best of our knowledge, no detailed computational study on this dye is available. Computational study gives insight into the geometric and electronic structure of the dye sensitizer and serves as the starting point for rational design of new dyes with desirable properties such as improved spectral coverage. We also investigate rational design through chemical modifications on the structure of this dye sensitizer aiming at red-shifting and broadening the absorption spectra of the S9 dye that might enhance the efficiency of DSSC by utilising a greater fraction of the solar spectrum.

## 2. Methods and computational details

The structure of the Carbz-PAHTDDT (S9) dye in three dimensional (3D) spaces was obtained through geometry optimizations in vacuum and in dichloromethane (DCM, $CH_2Cl_2$) solution, respectively. Density functional theory (DFT) based PBE0 hybrid density functionals [28] and polarized split-valence triple-zeta 6-311G(d) basis set, that is, the PBE0/6-311G(d) model, was employed in the calculations without any symmetry restrictions. No imaginary frequencies were found on the optimized structure, which ensures that optimized structure of S9 dye is a true minimum structure. All *ab initio* calculations were performed in Gaussian09 package [29].

To analyse the charge population of the dye sensitizer, natural bond orbital (NBO) analysis was performed on the optimized structure in vacuum using the NBO 3.1 program [30] embedded into Gaussian09 package. The computations of the hyperpolarizability and NBO were carried out using the same PBE0/6-311G(d) model in vacuum. The solvent effects (i.e. DCM) on the absorption spectra and molecular energy levels are calculated using the polarizable conductor calculation method (CPCM) [31, 32].

Three hybrid DFT functionals, namely, B3LYP, PBE0 and BHandH with the same basis set, i.e., B3LYP/6-311G(d)//PBE0/6-311G(d), PBE0/6-311G(d)//PBE0/6-311G(d), and BHandH/6-311G(d)//PBE0/6-311G(d) models were employed for single point energy calculations in DCM solution to construct the molecular energy levels and isodensity plots.

To accurately reproduce the experimental photo physical results, such as $\lambda_{max}$ of S9 dye sensitizer, several standard hybrid DFT functional (i.e., B3LYP [33], PBE0 [28] and BHandH [34]) and long-range corrected (LC) DFT functionals (such as CAM-

B3LYP [35], ωB97XD [36] and LC-ωPBE [37-39]) have been included for the TDDFT calculations. In the standard hybrid DFT functionals, the Hartree-Fock (HF) exchange energy component $V_x$ in the exchange-correlation $V_{xc}$ term increases from 20% (HF) in B3LYP to 25% HF in PBE0, and to 50% HF in BHandH (half-and-half functional which is implemented in Gaussian09 with 50% HF [40]). The LC DFT functionals with a damping parameter are employed in the present study. For example, the CAM-B3LYP functional consists of 65% of $V_x$ (HF) and 35% of $V_c$ (B88) at long-range, whereas 19% of $V_x$ (HF) and 81% of $V_c$ (B88) at short-range with a damping parameter of $\omega = 0.33\ a_0^{-1}$ [35]. In the ωB97XD functional, the short-range HF (exact) exchange is 22.20% and the damping parameter ω is $0.2\ a_0^{-1}$ [36]. The third LC-ωPBE functional uses $\omega = 0.4\ a_0^{-1}$ and no short-range exchange [37].

We have also investigated two new dyes which are rationally designed by structural changes in the π-conjugated bridge of the reference Carbz-PAHTDDT (S9) dye. Both new dyes are computationally studied using the same method of the S9 dye study.

## 3. Results and discussion

### 3.1. Geometrical structures and design of the new dyes

The backbone structure of the Carbz-PAHTDTT dye (S9) is given in Figure 1(a). This dye exhibits an electron-rich donor group (D), a π-conjugated bridge or linker and an acceptor moiety (A) as marked in the figure by three boxes. As a result, S9 is a D-π-A dye which is a common structure for organic dye sensitizers [16, 41-46]. A two-carbazole-unit substituted triphenylamine group is employed as the electron donor unit (D) of the dye. It was previously shown that this donor structure suppresses the close π-stacked aggregation between the donor moieties of dye sensitizers adsorbed onto the surface of $TiO_2$ semiconductor [47]. Aggregation can result in intermolecular quenching and also leads to dye molecules which are not functionally attached to the semiconductor's surface and work like filters [48]. This phenomenon is known to be a detrimental factor of the efficiency for DSSC which should be avoided either by structural design or by employing co-adsorbents [49]. The non-coplanar structure of the electron donating moiety (D) can enhance thermal stability of dye sensitizer molecules by decreasing the contact between them. Thermal stability of dye sensitizer is an important factor for long term stability of functional solar cells [47].

The π-bridge (linker, the middle box in Figure 1(a)) consists of five pentagon rings which are labelled as I, II, III, IV and V in Figure 1(a). A dithienothiophene (DTT) unit forms central part of the π-conjugated bridge of the S9 dye. This moiety leads to a better stability of the dye sensitizer in high polarity electrolytes used in DSSC. To provide additional double conjugation into the linker moiety [50] , two hexanyl ($C_6H_{13}$) chain-substituted thiophene rings (i.e. 3-hexylthiophene or rings I and V in Figure 1(a)) exist in the π-conjugated bridge of the S9 dye which can form either trans or cis isomers. A *cis*-S9 is formed when both of the hexanyl chains ($C_6H_{13}$) are in the same side of the π-bridge, or a *trans*-S9 isomer is formed if the hexanyl chain ($C_6H_{13}$) groups locate in the different sides of the π-bridge. The long hexanyl chains suppress the aggregation of the dye molecules, and also enable longer electron life time ($\tau$) [51]. The present calculations indicate that the *cis*-S9 isomer possesses a total energy of approximately 4.6 kJ·mol$^{-1}$ less than the total energy of the trans conformer, indicating that the S9 dye slightly favours the cis conformation. Therefore, only cis conformation will be studied throughout this work. On the acceptor side of this dye (A), the conventional acceptor moiety is employed which contains the cyano group as an electron withdrawing group and the carboxyl group as an anchoring unit to attach the dye onto the TiO$_2$ semiconductor.

As learned from our previous study on rational design of D-π-A dyes [25], all three moieties of a dye, i.e., the donor, the π-bridge and the acceptor can be modified to produce new dyes. For the purpose of the rational design in this study, the π-bridge of the S9 dye has been modified to produce a pair of new dyes. In the π-conjugated bridge of this dye, a dithienothiophene unit (DTT) is employed. Kwon *et al.* who synthesized S9 dye, have also reported another DTT-based dye sensitizer (DAHTDTT 13) with a similar structure to S9 dye which only differs in its D group [50]. The absorption spectra of these two dye sensitizers are very similar for the visible portion of the spectrum, i.e., λ>400 nm (refer to Figure S1 in supporting information). As a result, in the present study, instead of making changes in the D-group and A-group, the linker (i.e., the π-bridge) of the S9 dye is modified.

An aim of the design of the new dyes is to extend the absorption spectra to near infrared (NIR) region by reducing the HOMO-LUMO energy gap of the dye sensitizer. As a result, two new derivatives dyes (S9-D1 and S9-D2) are designed from the original (*cis*-) Carbz-PAHTDTT (S9) dye through the modification of the π-bridge linker. In S9-D1 (Figure 1(c)), the $X_1$ and $X_2$ groups in S9 dye (Figure 1(a)) were replaced by the N= groups, respectively, but in S9-D2 dye (Figure 1(d)), the $X_1$ and $X_2$ groups in Figure 1(a) were substituted by the -NH groups, respectively. The obtained π-bridges of the optimized structures Carbz-PAHTDTT (S9), S9-D1 and S9-D2 are given in Figure 1(b), Figure 1(c) and Figure 1(d), respectively.

The new dyes will exhibit differences in electronic properties as they are different compounds. Atomic charges according to the natural bond orbital (NBO) scheme of the π-conjugated bridges of the dyes are also given in Figure 1(b)-(d) for three dyes S9, S9-D1 and S9-D2, respectively. It is not a surprise that the atomic charges change more apparently at the positions local to the $X_1$ and $X_2$ atoms in the new dyes, whereas only small changes in atoms away from $X_1$ and $X_2$ are observed. For example, atoms $C_{(45)}$ and $C_{(48)}$ directly bond with $X_1$, whereas $C_{(49)}$ and $C_{(52)}$ directly bond with $X_2$. The atomic charges for these sites in S9 are negative, i.e., $X_1$=S: -0.185 a.u. for $C_{(45)}$ and -0.213 a.u. for $C_{(48)}$; and $X_2$=S: -0.186 a.u. for $C_{(49)}$ and -0.246 a.u. for $C_{(52)}$. However, the atomic charges in these sites in the derivatives, S9-D1 and S9-D2, switch their signs from being negative in S9 to being positive in S9-D1 and S9-D2, regardless the substituted groups are electron donating (-NH-) or electron withdrawing (-N-) groups.

The total NBO charges of the π-conjugated bridge (linker) in the D-π-A dyes can be either negative or positive. However, the overall net charge for the donor section (D) of the dyes is always positive, whereas the acceptor section (A) of the dyes is always negative. Although in the new dyes similar trend exist in the changes of the NBO atomic charges locally, the total NBO charges over the π-conjugated bridge (linker) of the dyes are not the same. Such the total charges over the π-conjugated bridge (linker) are calculated at +0.062 a.u., -0.029 a.u. and +0.118 a.u., respectively, for S9, S9-D1 and S9-D2. The fact that the original S9 dye and the new dye S9-D2 possess positive charges of the linker suggests that π-conjugated bridges of these dyes exhibit electron-donating character. On the contrary, the negatively charged linker of the S9-D1 dye shows that the chemical modifications with the –NH- group alter the electron-donating character of the linker in the original dye (S9) to an electron-withdrawing character of the linker in the S9-D1 dye.

Table 1 lists the important molecular properties of the Carbz-PAHTDTT (S9) dye (*cis* isomer) and the new S9-D1 and S9-D2 dyes obtained in vacuum. As all the dyes are either new dyes (S9-D1 and S9-D2) or recently synthesised dye (S9), only very limited information is available for comparison. However, the models such as the PBE0/6-311G(d) model have been shown to be reliable in previous studies [25, 28, 52]. The π-conjugated bridge length ($L_\pi$) of the D-π-A dye, defined as the direct distance between $C_{(43)}$ and $C_{(61)}$, is calculated at 17.14 Å for S9 dye which is shortened in S9-D1 (16.33 Å) and S9-D2 (16.52 Å) as the N atoms (S9-D1 and S9-D2) have smaller radius than S atoms (S9), which is also reflected by the calculated molecular sizes (i.e., the electronic spatial extent $<R^2>$) of the dyes. The dipole moments (μ) of the S9-D2 dye (μ=5.12 Debye) exhibit very similar values to the original S9 dye (μ=5.10 Debye), whereas the dipole moment of S9-D1 (6.72 Debye) exhibits larger changes.

### 3.2. Electronic and optical properties

Based on the cyclic voltammetry measurements of the Carbz-PAHTDTT (S9) dye in $CH_2Cl_2$ (DCM) solution reported in the supplementary information of the reference paper [13], the experimental energy values for the HOMO, LUMO and HOMO-LUMO gap are quantified at -5.08 eV, -2.97 eV and 2.11 eV, respectively. In the present study, three hybrid functionals (B3LYP, PBE0 and BHandH) have been employed to calculate the frontier molecular orbital energies of the S9 dye in DCM solution. These three functionals have an increasing trend in the percentage of HF (exact) exchange as B3LYP (20%) < PBE0 (25%) < BHandH (50%). The calculations show that the B3LYP functional provides the best agreement with the experiment by less than 0.02 eV deviations from the experimental values, indicating that exchange energy is less important than correlation energy here. Thus the B3LYP/6-311G(d) model is employed to calculate the energy levels of rationally-designed dyes S9-D1 and S9-D2 (refer to Table S1 in supporting information for more details).

Figure 2 compares the calculated frontier molecular orbital energy levels of the S9 dye and the new dyes S9-D1 and S9-D2 in DCM solution, focusing on the HOMO–LUMO energy gap. As seen in this figure, the HOMO-LUMO gap of the dyes reduces from 2.08 eV in S9 to 1.66 eV in S9-D1 and 1.88 eV in S9-D2. However, the HOMO-LUMO gap reductions in S9-D1 and S9-D2 are very different: in S9-D1, the most significant change is the reduction of the LUMO energy, from -2.99 eV in S9 to -3.66 eV in S9-D1, although the energy of the HOMO of S9-D1 also exhibits a small reduction, from -5.08 eV (S9) to -5.32 eV (S9-D1). On the other hand, the HOMO-LUMO gap reduction in S9-D2 dye is achieved by shifting up the energy level of the HOMO, from -5.08 eV to -4.79 eV, whereas the energy of the LUMO almost remains the same as -2.99 eV in S9 and -2.91 eV in S9-D2. The HOMO-LUMO energy gap reductions of the new dyes in Figure 2 are also seen in the corresponding orbitals of the dyes. When the orbitals are more localised, the energies increase, whereas when the orbitals are more delocalised, the energies decrease.

The new dye S9-D1 has similar HOMO distribution to the reference structure but with more contribution from the carbazole-units in the donor moiety and the second hexanyl chain-substituted thiophene ring (i.e. ring V) in the linker moiety. However, the LUMO of S9-D1 is extended into a phenyl group in the donor moiety of the dye structure and a small density decrement is observed on the carboxyl group (A section) which might lead to a weaker electron coupling with semiconductor surface compared to the reference S9 structure. Also, S9-D2 dye shows very similar HOMO and LUMO distribution to the reference S9 dye with the HOMO being slightly shifted from the donor section toward the π-conjugated bridge.

It is well-known that an important feature of organic D-π-A dye sensitizers is intra-molecular charge transfer character (ICT). Edvinsson *et al.* [54] studied some perylene-based sensitizers for the relationship between ICT character of the molecules and their performance in DSSC. It was found that the photocurrent and the overall solar-to-electrical energy conversion efficiency improve remarkably with increasing ICT character of the dyes. A later study of Tian and co-workers on triphenylamine dyes [55] suggests that an effective ICT has a positive effect on the performance of DSSC. Since first hyperpolarizability ($\beta$) of a push-pull organic dye is associated with the ICT character [56, 57], the total hyperpolarizability ($\beta_{tot}$) [58] of the Carbz-PAHTDTT (S9) and the new S9-D1 and S9-D2 dyes is given by:

$$\beta_{tot}= [ (\beta_{xxx} + \beta_{xyy} + \beta_{xzz})^2 + ( \beta_{yyy} + \beta_{yzz} + \beta_{yxx})^2 + ( \beta_{zzz} + \beta_{zxx} + \beta_{zyy})^2 ]^{1/2} \quad (1)$$

where the tensor components, i.e. $\beta_{xxx}$, $\beta_{xxy}$, $\beta_{xyy}$, $\beta_{yyy}$, $\beta_{xxz}$, $\beta_{xyz}$, $\beta_{yyz}$, $\beta_{xzz}$, $\beta_{yzz}$, and $\beta_{zzz}$ are given in Table S2 in supporting information. The $\beta_{tot}$ are calculated as 804.802 esu for S9, 2190.035 esu for S9-D1 and 856.034 esu for S9-D2. Both of the new dyes S9-D1 and S9-D2 show the $\beta_{tot}$ values higher than the reference dye and the $\beta_{tot}$ of the S9-D1 is significantly larger than the S9 dye. Our results together with other studies show that HOMO-LUMO gap is a critical factor in determining the $\beta_{tot}$ value [56, 58]. The S9-D1 dye exhibits both significantly larger $\beta_{tot}$ value as well as reduced HOMO-LUMO gap which are desirable properties for good performing dye sensitizers for the application in DSSC.

## 3.3. Excitation energies and UV-Vis spectra

The experimental absorption spectrum of the S9 dye was measured in the dichloride methane (DCM) solution [13]. The three main absorption bands in the UV-Vis spectral region of 300-800 nm of the S9 dye are reported at $\lambda_1$=491 nm, $\lambda_2$=426 nm, and $\lambda_3$=330 nm [13]. Table 2 compares the experimental measurement with theoretical results using TDDFT with respect to different DFT models indicated in the previous section for the original S9 dye. To assess the overall performance of TDDFT functionals with reference to the experimental values in this table, the mean absolute error (MAE) criterion is employed as,

$$\text{MAE}= 1/n \sum_{i=1}^{n} \left|\lambda_i^{calc.} - \lambda_i^{expt.}\right|, (n=3) \tag{2}$$

The MAEs of the DFT functionals are given in the last row of Table 2. In the table, the CAM-B3LYP model and the BHandH are compatible with MAEs being 14 nm and 18 nm, respectively. Next comes the ωB97XD (MAE=23 nm) and the LC-ωPBE (MAE=52 nm) models. The PBE0 (MAE=108 nm) and B3LYP (MAE=149 nm) models exhibit the least accurate performance on prediction of the spectral line positions. Three long-range (LC) corrected DFT functionals, namely CAM-B3LYP, ωB97XD and LC-ωPBE show a good general performance in reproducing the experimental main bands. Non-LC hybrid functionals are not usually suitable and accurate at large distances for electron excitations to high orbitals. That is because the non-Coulomb fraction of exchange functionals usually diminishes very fast. In the range-separated (or LC) hybrid DFT functionals, the Coulomb energy is divided into long-range and short-range energies. The HF exchange interaction is included in the long-range part and the DFT exchange interaction is included in the short-range part [59]. The ω (in bohr$^{-1}$) parameter is a damping parameter which controls the range of the inter-electronic separation between these two terms. The present results are in good agreement with other studies [60-65], suggesting that long-range corrected CAM-B3LYP functional can be a good model in the TDDFT calculations of large-sized organic dye molecules with a spatially extended π system, as the charge transfer transitions take place through space in such dyes.

The results in Table 2 also suggest that without long-range corrections, the inclusion of the Hartree-Fock (HF) exchange energy is important to reproduce the major band, $\lambda_1$. The TD-BHandH DFT model gives the major absorption peaks of the S9 dye the most accurate results and the TD-B3LYP DFT model produces the least accurate results in the table. For example, the TD-BHandH DFT model almost reproduces the spectral line position of the major absorption band at $\lambda_1$= 490 nm with respect to the experiment at $\lambda_1$=491 nm. This model also closely reproduces the other minor bands at $\lambda_2$=402 nm (expt. 426 nm) and $\lambda_2$=360 nm (expt. 330 nm). Without sufficient inclusion of the exchange energy in the DFT functionals, the B3LYP and PBE0 hybrid functionals are unable to produce spectral band positions with sufficient accuracy as seen previously [66, 67].

Table 2 also collects the effect of modifications on shifting the spectral peaks. For S9-D1, a remarkable bathochromic shift (i.e. to the longer wavelengths or red-shift) of 172 nm on $\lambda_1$ compared to the reference S9 dye is observed. This band is mainly composed of an excitation transition from HOMO → LUMO both for the reference S9 dye and for the new S9-D1 structure (refer to Table S3 in supporting information for detailed assignment of the transitions). Figure 3 reports the simulated UV-vis spectra of the three dyes, S9, S9-D1 and S9-D2. As discussed earlier, the energy gap between HOMO and LUMO of S9-D1 is significantly reduced compared to that of S9 dye, which in turn results in the red-shift of $\lambda_1$ as seen in this figure. Very large red-shift of $\lambda_2$ and $\lambda_3$ are also observed for S9-D1 compared to the S9 dye. The $\lambda_1$ band of the new dye S9-D1 indeed outperforms the original Carbz-PAHTDTT (S9) dye with not only a significant preferred spectral shift on the position of this band, but also this band covers a broader region with nearly doubled full width at half maximum (FWHM) than the original S9 dye. The significant bathochromic shift and broadening of the UV-Vis spectra of S9-D1 structure indicates its enhanced light harvesting capability which is an important criterion for a well-performing dye sensitizer employed in DSSC. In the S9-D2 dye, the $\lambda_1$ and $\lambda_3$ spectral bands both show a preferred bathochromic shift of 44 nm compared to the S9 dye. However, an unwanted hypsochromic shift (i.e. to the shorter wavelengths or blue-shift) of -32 nm on the $\lambda_2$ spectral band was calculated for this (S9-D2) dye.

**4. Conclusions**

The recently published Carbz-PAHTDTT (S9) organic dye sensitizer was studied theoretically using a number of DFT models in vacuum and in DCM solution. Good agreement with experimental results indicate that the B3LYP/6-311G(d)//PBE0/6-311G(d) model in solution reproduced the energy levels of HOMO and LUMO more accurately than other models in this study. However, to produce good agreement with the experimental UV-Vis spectra of the S9 dye, long-range corrected functionals, such as CAM-B3LYP and half-and-half functionals such as BHandH functional need to be considered.

In the present study, a pair of new dyes, S9-D1 and S9-D2, was designed through chemical modifications of the π-conjugated bridge of the reference S9 dye. These new dyes showed a reduced HOMO-LUMO energy gap compared to that of the S9 dye, through lowering the LUMO energy (S9-D1) or lifting up the HOMO energy (S9-

D2). A significant red-shift and broadening of the resulting absorption spectrum of the S9-D1 is achieved. The present study explored a useful direction of rational design for new dyes in DSSC.

**Acknowledgements**

NM would like to thank Swinburne University Vice-chancellor's Postgraduate Award. Dr. Tae-Hyuk Kwon must be acknowledged for helpful discussions. NM and FW thank Victorian Partnership for Advanced Computing (VPAC) and Swinburne University Supercomputing Facilities for computer resources.

**Table 1:** The selected bond length, dihedrals, π-lengths[a] and dipole moment of the S9, S9-D1 and S9-D2 dyes*.

|  | S9 | S9-D1 | S9-D2 |
|---|---|---|---|
| $L_\pi$ [a] (Å) | 17.14 | 16.33 | 16.52 |
| $C_{48}$-$C_{49}$ (Å) | 1.44 | 1.37 | 1.44 |
| $X_1$-$C_{48}$-$C_{49}$-$X_2$ (°) | -144.91 | -179.09 | -157.07 |
| $X_2$-$C_{52}$-$C_{53}$-$S_4$ (°) | 0.93 | -0.79 | -0.10 |
| $S_4$-$C_{56}$-$C_{57}$-$S_5$ (°) | 150.10 | 157.57 | 148.19 |
| $<R^2>$ (a.u) | 275867.82 | 256546.52 | 260606.50 |
| μ (Debye) | 5.10 | 6.72 | 5.12 |

*Optimized at PBE0/6-311G(d) level.

[a] Direct distance of $C_{(43)}$-$C_{(61)}$.

**Table 2: The most intense absorption peaks of the S9, S9-D1 and S9-D2 dyes.**

| Method[a] | Carbz-PAHTDTT (S9) | | | | | | | S9-D1 | | S9-D2 | |
|---|---|---|---|---|---|---|---|---|---|---|---|
| | TD-B3LYP | TD-PBE0 | TD-LC-ωPBE | TD-ωB97XD | TD-CAM-B3LYP | TD-BHandH | Exp.[b] | TD-BHandH | $\Delta\lambda$[c] | TD-BHandH | $\Delta\lambda$[d] |
| $\lambda_1$(nm) | 668 | 611 | 420 | 460 | 479 | 490 | 491 | 662 | 172 | 535 | 45 |
| $\lambda_2$(nm) | 540 | 498 | 365 | 391 | 401 | 402 | 426 | 528 | 126 | 394 | -8 |
| $\lambda_3$(nm) | 488 | 463 | 306 | 325 | 335 | 360 | 330 | 440 | 80 | 374 | 14 |
| MAE[e] | 149 | 108 | 52 | 23 | 14 | 18 | | | | | |

(a) All TDDFT calculations are performed in DCM solution using CPCM solvation model on geometries optimized at CPCM -PBE0/6-311G(d).

(b) See supplementary information of [13].

(c) $\Delta\lambda = \lambda(S9\text{-}D1) - \lambda(S9)$, method= TD-BHandH.

(d) $\Delta\lambda = \lambda(S9\text{-}D2) - \lambda(S9)$, method= TD-BHandH.

(e) MAE= $1/n \sum_{i=1}^{n} |\lambda_i^{calc.} - \lambda_i^{expt.}|$, (n=3)

**Figure Captions**

**Figure 1.** Sketch of the reference S9 dye and the structure of the π-conjugated bridge of S9, S9-D1 and S9-D2 dyes showing NBO charge of atoms in the linker. Note that hexanyl chains are not included.

**Figure 2.** Calculated frontier MO energy levels using B3LYP/6-311G(d) // PBE0/6-311G(d) model in DCM solution and isodensity surfaces of the HOMO and LUMO of S9, S9-D1 and S9-D2 dyes.

**Figure 3.** The simulated UV-Vis spectra of three dyes, S9, S9-D1 and S9-D2 using TD-BHandH/6-311G(d) model in DCM solution.

Fig.1

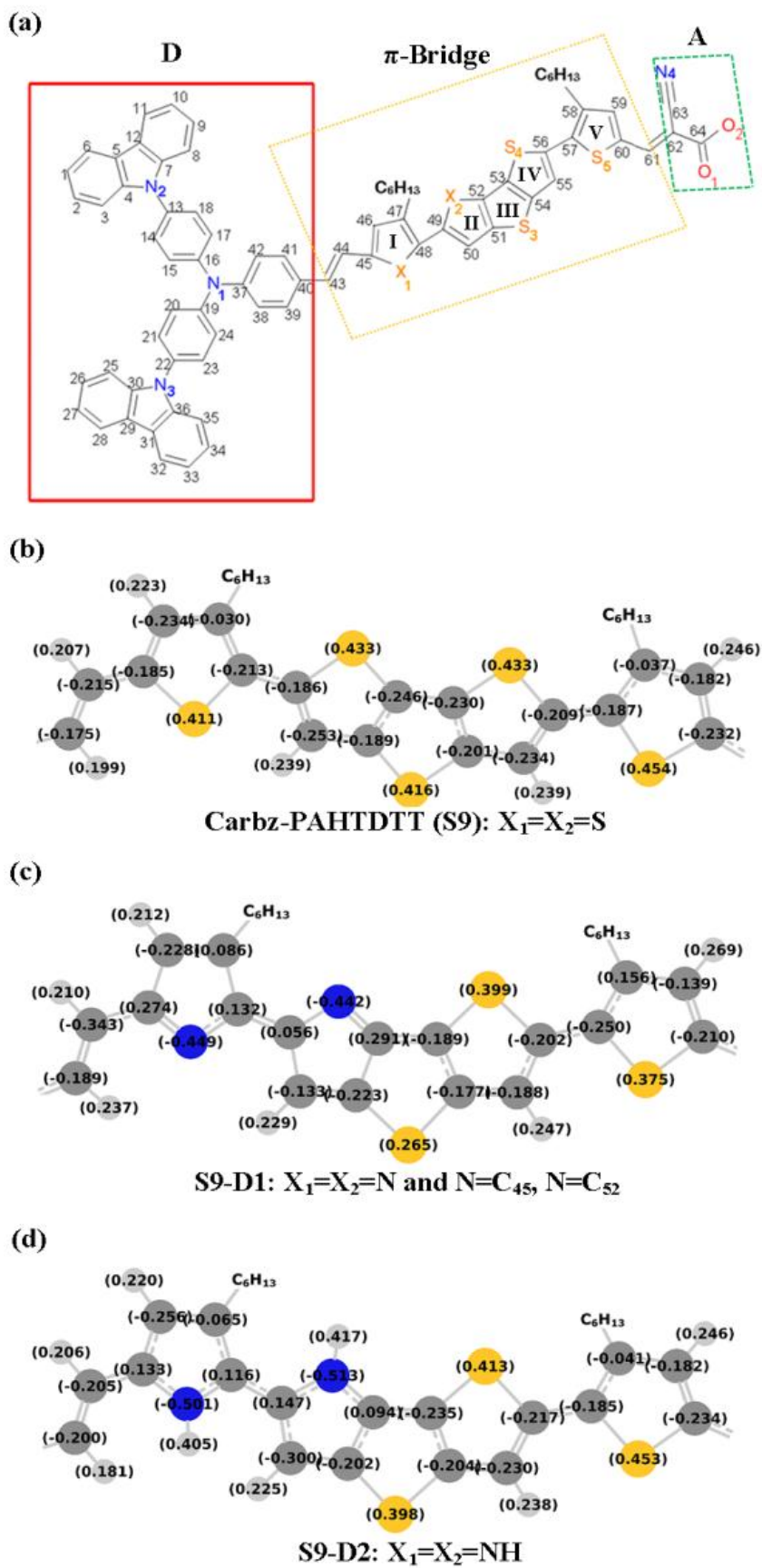

Fig. 2

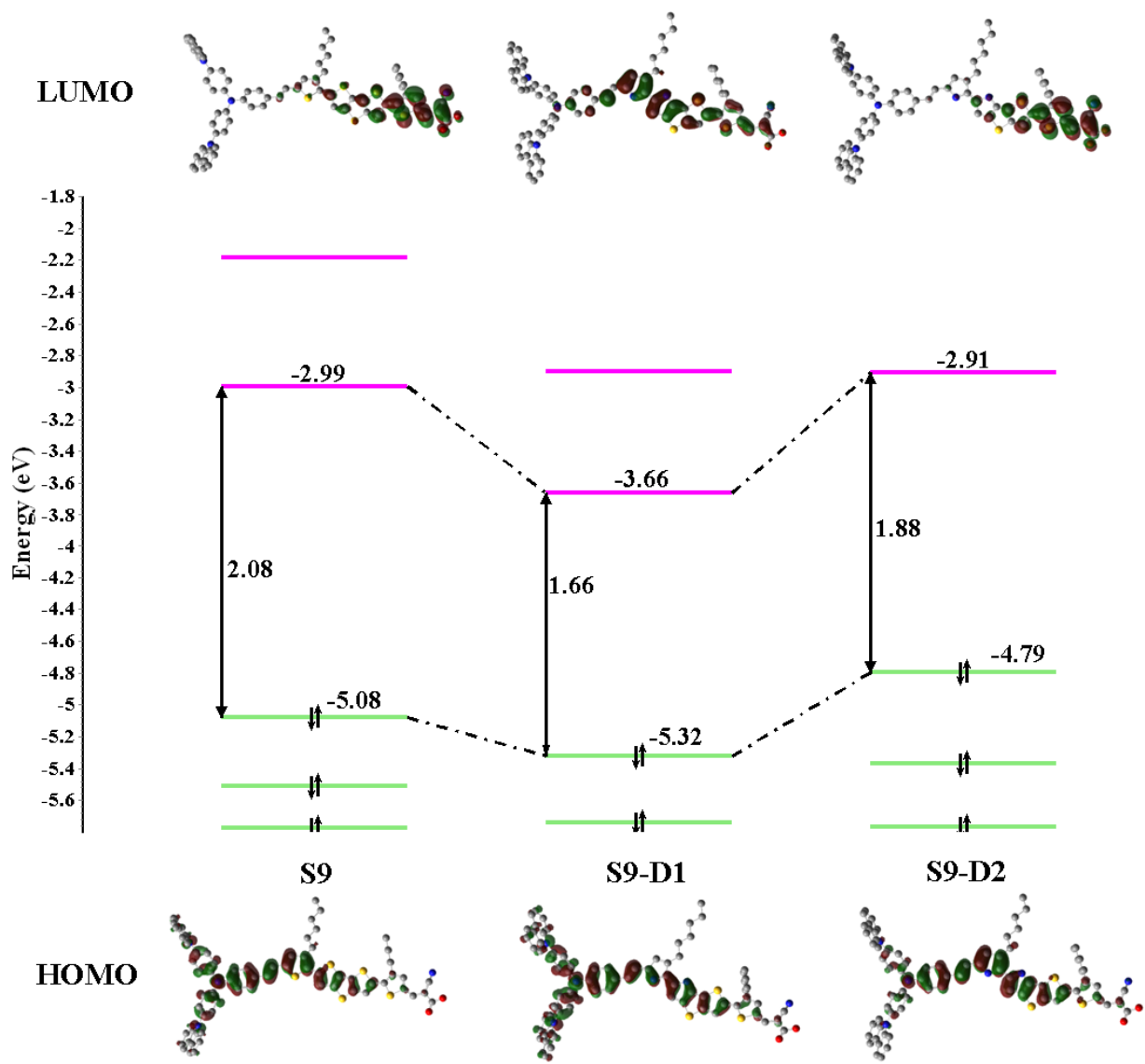

Fig. 3

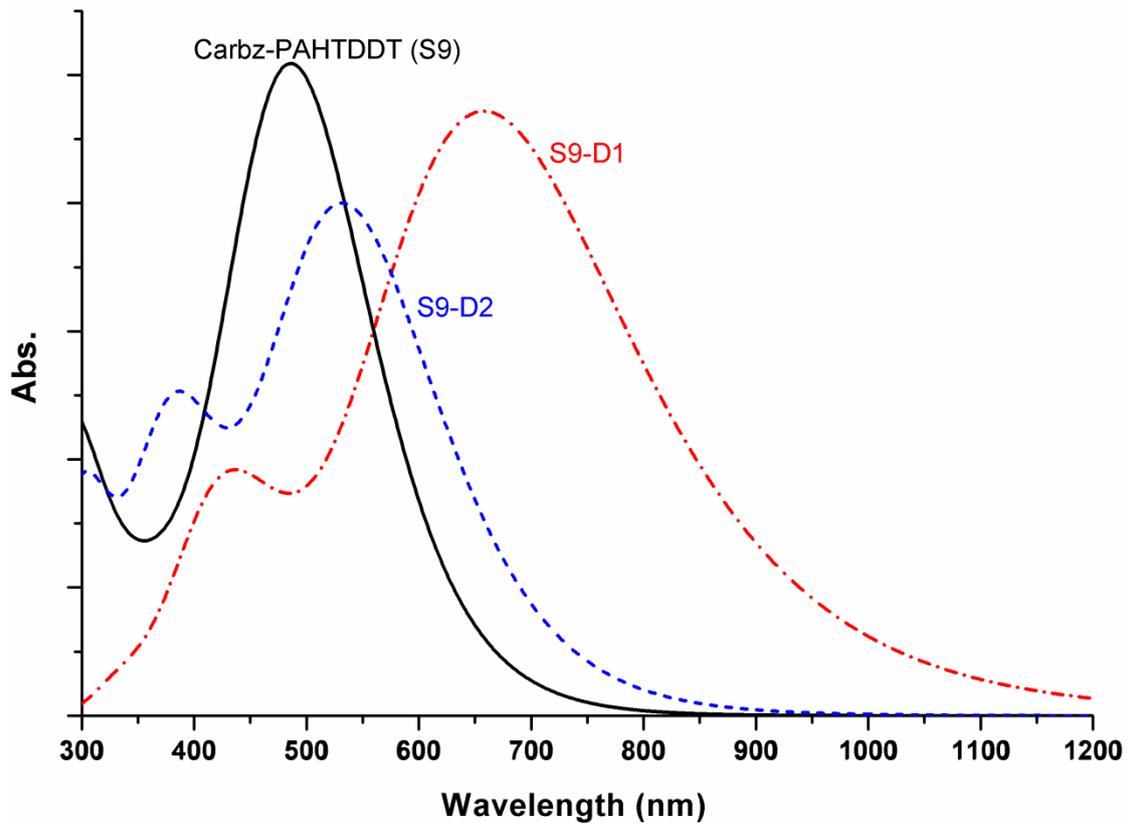